\documentclass[twocolumn,prl,showpacs,amsmath,amstex,amssymb,mathfonts]{revtex4-1}
\pdfoutput=1
\usepackage{amsthm,color,amsfonts,graphicx,verbatim}
\usepackage{amsmath}
\usepackage{amssymb}
\usepackage{amsthm}
\usepackage{amsfonts}
\usepackage{listings}
\usepackage{enumerate}
\usepackage{latexsym}
\usepackage{psfrag}
\usepackage{bm}
\usepackage{graphicx}
\usepackage{dsfont}

\usepackage{hyperref}
\hypersetup{
    bookmarks=true,         
    unicode=false,          
    pdftoolbar=true,        
    pdfmenubar=true,        
    pdffitwindow=false,     
    pdfstartview={FitH},    
    pdftitle={Many-body localization in periodically driven systems},    
    pdfauthor={},     
    pdfsubject={},   
    pdfcreator={},   
    pdfproducer={}, 
    pdfkeywords={keyword1} {key2} {key3}, 
    pdfnewwindow=true,      
    colorlinks=true,       
    linkcolor=black,          
    citecolor=blue,        
    filecolor=magenta,      
    urlcolor=blue           
}

\newcommand{\be}{\begin{equation}}
\newcommand{\ee}{\end{equation}}
\newcommand{\bea}{\begin{eqnarray}}
\newcommand{\eea}{\end{eqnarray}}

\newcommand{\la}{\langle}
\newcommand{\ra}{\rangle}

\renewcommand{\phi}{\varphi}
\renewcommand{\epsilon}{\varepsilon}

\begin{document}
\title{Many-body localization in periodically driven systems}

\author{Pedro Ponte,$^{1,2}$ Z. Papi\'c,$^{1,3}$ Fran\c{c}ois Huveneers$^4$, and Dmitry A. Abanin$^{1,3}$}

\affiliation{$^1$ Perimeter Institute for Theoretical Physics, Waterloo, ON N2L 2Y5, Canada}
\affiliation{$^2$ Department of Physics and Astronomy, University of Waterloo, Ontario, N2L 3G1, Canada}
\affiliation{$^3$ Institute for Quantum Computing, Waterloo, ON N2L 3G1, Canada}
\affiliation{$^4$ CEREMADE, Universite Paris-Dauphine, France}

\date{\today}
\begin{abstract}

We consider disordered many-body systems with periodic time-dependent Hamiltonians in one spatial dimension. By studying the properties of the Floquet eigenstates, we identify two distinct phases: (i) a many-body localized (MBL) phase, in which almost all eigenstates have area-law entanglement entropy, and the eigenstate thermalization hypothesis (ETH) is violated, and (ii) a delocalized phase, in which eigenstates have volume-law entanglement and obey the ETH. MBL phase exhibits logarithmic in time growth of entanglement entropy for initial product states, which distinguishes it from the delocalized phase. We propose an effective model of the MBL phase in terms of an extensive number of emergent local integrals of motion (LIOM), which naturally explains the spectral and dynamical properties of this phase. Numerical data, obtained by exact diagonalization and time-evolving block decimation methods, suggests a direct transition between the two phases. Our results show that many-body localization is not destroyed by sufficiently weak periodic driving. 

\end{abstract}

\maketitle

{\bf Introduction.} The dynamics of closed quantum many-body systems driven out of equilibrium has been the subject of intense investigation over the past decade~\cite{Polkovnikov11,Bloch08}. Many-body systems with local Hamiltonians broadly fall into two classes with distinct dynamical properties: ergodic systems, which reach local thermal equilibrium as a result of the Hamiltonian evolution, and non-ergodic ones which fail to thermalize. Thermalization in isolated ergodic systems can be linked to the properties of individual many-body eigenstates that are locally thermal~\cite{deutsch, srednicki,Rigol08}.

While a complete classification of non-ergodic systems remains an open problem, it has recently been established that many-body localization~\cite{Basko06, Mirlin05, Oganesyan07, Pal10, Serbyn13-1, Serbyn13-2, Huse13,Moore12,Prosen08,Vosk13, bauer,Pekker14} provides a robust mechanism of ergodicity breaking in systems with quenched disorder. Many-body localized (MBL) systems are characterized by an extensive number of quasi-local conservation laws~\cite{Serbyn13-2,Huse13}, which strongly restrict quantum dynamics and prevent energy transport and thermalization. MBL systems have universal dynamical properties, such as the logarithmic-in-time growth of entanglement entropy for initial product states~\cite{Prosen08, Moore12, Serbyn13-1,Serbyn13-2,Huse13,Vosk13}, in contrast to ergodic and Bethe-ansatz-integrable systems where entanglement spreads linearly in time~\cite{Calabrese05,Chiara05,Kim13}. 

In this paper, we study disordered many-body systems with local time-dependent Hamiltonians $H(t)$ that vary periodically in time, $H(t+T)=H(t)$. The properties of periodically driven systems are determined by the unitary Floquet operator $\hat F$, i.e., the evolution operator over one period: 
\be\label{eq:Floquet}
\hat{F}={\cal T}\exp\{-i\int_0^T  H(t) dt\}, 
\ee 
where ${\cal T}\exp$ denotes a time-ordered exponential. In the eigenstate basis $|\psi_\alpha\ra$, $\hat F$ takes the form
$
\hat F=\sum_{\alpha=1}^{\mathcal{D}} e^{-i\theta_\alpha}|\psi_\alpha\ra \la \psi_\alpha|, 
$
where $\mathcal{D}$ is the Hilbert space dimension, and the quasi-energies $\theta_\alpha$ can be chosen to lie in the interval $[0;2\pi)$. One can introduce an effective Floquet Hamiltonian $H_F$ as $\hat F=e^{-iH_F}$, with eigenstates $|\psi_\alpha\ra$ and eigenvalues $\theta_\alpha+2\pi n_\alpha$, where $n_\alpha$ is an arbitrary integer. Generally it is not known whether there exists a choice of $n_\alpha$ which brings $H_F$ into a sum of local terms, and therefore the Floquet problem cannot be reduced to the study of time-independent local Hamiltonians. It is the goal of this paper to explore different regimes of the Floquet dynamics in disordered many-body systems. 
 
We consider a generic class of periodically driven 1D models with quenched disorder, and find that, as system's parameters are varied, two distinct phases are realized, which differ in the structure of their Floquet eigenstates, as well as in dynamical properties. One of them is the MBL phase in which the Floquet eigenstates at arbitrary quasi-energy obey the area-law for entanglement entropy, similar to the ground states in gapped systems. Level repulsion is absent, and the statistics of quasi-energy levels follows the Poisson statistics. Further, in the limit of an infinite system, the eigenstates with similar quasi-energies typically have different local properties, thus the eigenstate thermalization hypothesis (ETH)~\cite{deutsch, srednicki,Rigol08} breaks down. The second phase is the delocalized (ergodic) phase. Here the Floquet eigenstates have an extensive, volume-law entanglement; the quasi-energy levels repel, and their statistics is described by Circular Orthogonal Ensemble (COE). ETH holds in this phase, and the Floquet eigenstates have identical local properties, described by an infinite-temperature Gibbs ensemble. 

The two phases can furthermore be distinguished by their dynamical properties, e.g., the time evolution of the system prepared in a product state, which can be efficiently simulated numerically. In the MBL phase, the states retain local memory of the initial state, and local observables at long times are correlated with their initial values. Similar to MBL systems with time-independent Hamiltonians, entanglement entropy grows logarithmically in time. This behavior reflects the presence of emergent local integrals of motion~\cite{Serbyn13-2, Huse13}, which we explicitly construct following Ref.~\cite{Chandran14} (see also Ref.~\cite{Ros14}). In contrast, in the delocalized phase local observables relax to their ``equilibrium" values at long times, which are given by the infinite-temperature Gibbs ensemble. In this case, entanglement spreads much faster, and we find a behaviour consistent with the linear growth of entanglement. 

Our results complement previous works~\cite{Prosen98,Alessio13,Alessio14,Lazarides14}, which considered translationally invariant driven systems, as well as Ref.~\cite{Ponte14}, where the behaviour of disordered many-body systems under local driving was studied. 

{\bf Model.} Our system is a 1D spin 1/2 chain with open boundary conditions. Following Refs.~\cite{Prosen98,Alessio13}, we consider a driving protocol in which the system's Hamiltonian is periodically switched between two operators, $H_0$ and $H_1$, both of which are sums of local terms.  An example of a disordered Hamiltonian $H_0$, which describes an MBL phase and acts for time $T_0$, is
\be\label{eq:H}
H_0=\sum_i h_i \sigma_i^z+{J_z} \sigma_i^z \sigma_{i+1}^z, 
\ee
where random fields $h_i$ are uniformly distributed in the interval $[-W;W]$. The eigenstates of $H_0$ are product states. As a delocalizing Hamiltonian $H_1$ we choose 
\be\label{eq:kick}
H_1=J_x\sum_i \sigma_i^x\sigma_{i+1}^x+\sigma_i^y\sigma_{i+1}^y,
\ee
which acts for time $T_1$ such that the driving period is $T=T_0+T_1$. The associated Floquet operator is given by: 
\be\label{eq:Floquet_protocol}
\hat F=e^{-iH_0T_0} e^{-iH_1T_1}. 
\ee
The protocol  describes an MBL system periodically "kicked" with a delocalizing perturbation $H_1$, and can be viewed as a many-body generalization of a periodically kicked rotor model~\cite{casati,grempel,flmoore,lemarie,dittrich,izrailev,haake}. 
Recent work has argued that a similar protocol for translationally-invariant Hamiltonians results in an infinite-temperature state at long times, and therefore a non-local Floquet operator~\cite{Alessio14}. 
We fix $J_x=J_z=1/4$, $T_0=1$, $W=2.5$ and tune the strength of the kick -- $T_1$ -- observing a transition at critical $T_1^*$ between an MBL phase (small $T_1<T_1^*$) and an ergodic phase ($T_1>T_1^*$). We note that the model (\ref{eq:H},\ref{eq:kick}) always has one conserved quantity, the $z$-projection of the total spin, $S_z=\sum_{i=1}^N \sigma_i^z$. However, we have checked that this global conservation law is not essential for the existence of MBL and ergodic phases, by studying other models where $S_z$ is not conserved. 

\begin{figure}[htb]
\begin{center}
\includegraphics[width=0.9\columnwidth]{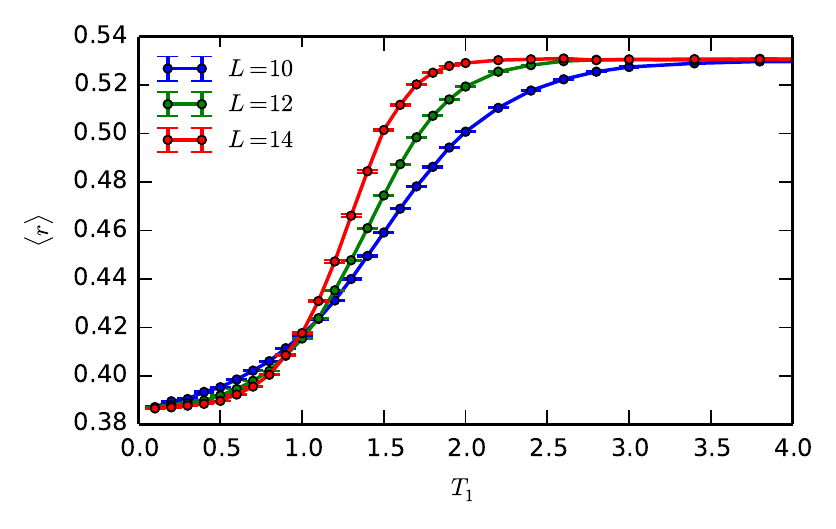}
\caption{ \label{Fig1} Disorder-averaged level statistics parameter $\la r\ra$ as a function of the ``kick" strength $T_1$. At small values of $T_1$, $\la r \ra\approx 0.386$, indicating Poisson statistics of quasi-energy levels (no level repulsion). At larger $T_1$ the system undergoes a transition into a delocalized phase with $\la r \ra\approx 0.53$, consistent with COE~\cite{Alessio14}. Data is for system sizes $L=10,12,14$, and averaging is performed over 1000 disorder realizations.}
\end{center}
\end{figure} 
{\bf Properties of Floquet eigenstates.} We first explore the properties of the Floquet eigenstates $|\psi_\alpha\ra$ and quasi-energy spectrum $\theta_\alpha \in [0;2\pi)$, using exact diagonalization (ED). By computing the consecutive quasi-energy gaps $\delta_\alpha=\theta_{\alpha+1}-\theta_\alpha$, we characterize the level statistics by their ratio 
$ r={\rm min} (\delta_\alpha, \delta_{\alpha+1})/{\rm max} (\delta_\alpha, \delta_{\alpha+1})$~\cite{Oganesyan07,Alessio14}. The averaged value of $r$ serves as a probe of ergodicity breaking: it allows one to distinguish between the Poisson and Wigner-Dyson level statistics. In Fig.~\ref{Fig1} we show $\la r\ra$ averaged over all quasi-energy spacings and over 1000 disorder realizations, for several system sizes. At small kick period $T_1$, $\la r \ra$ becomes increasingly close to the Poisson-statistics value $\la r \ra_{\rm POI}\approx 0.386$ as the system size is increased. This indicates the absence of level repulsion and suggests that ergodicity is broken at small $T_1$ and the system is in the MBL phase. At large $T_1$ parameter $\la r \ra$ is approximately equal to $0.53$, which is close to the COE value, $\la r\ra_{\rm COE}\approx 0.527$~\cite{Alessio14}. This suggests that at large $T_1$ the system delocalizes. The $\la r \ra$ curves for different system sizes cross at $T_1^*\approx 0.9$, suggesting a phase transition between MBL and ergodic phases. A drift of the crossing point towards smaller $T_1$ is observed, similar to the time-independent case~\cite{Pal10}. 

\begin{figure}[htb]
\begin{center}
\includegraphics[width=\columnwidth]{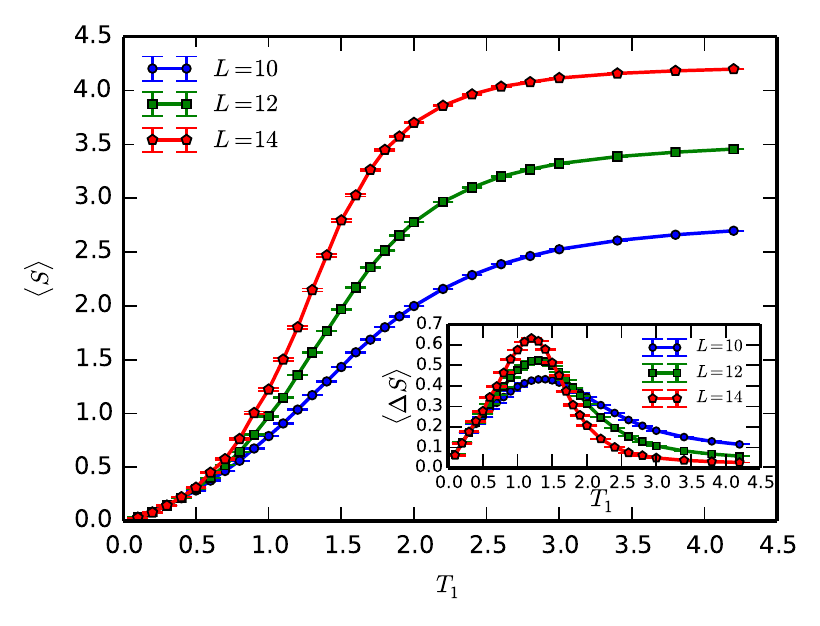}
\caption{ \label{Fig2} Averaged entanglement entropy $\la S \ra$  and its fluctuations $\la \Delta S \ra$ (inset) as a function of $T_1$. The scaling of entropy and its fluctuations with system size $L$ are consistent with the existence of an MBL and a delocalized phase for small and large $T_1$, respectively.}
\end{center}
\end{figure}

To further distinguish the nature of the two phases, we study the entanglement properties of the Floquet eigenstates. The expectation from the static case is that MBL eigenstates should obey an area law for entanglement entropy, i.e. in 1D their entropy should weakly depend on the chain size~\cite{Serbyn13-2,bauer}, while in the ergodic phase the eigenstates are thermal and their entropy scales as $L$. Fig.~\ref{Fig2} shows disorder- and ensemble-averaged von Neumann entropy $\la S \ra$ of the Floquet eigenstates, for the symmetric bipartition, plotted as a function of $T_1$. The markedly different scaling of $\la S \ra$ at small and large values of $T_1$ lends further support to the existence of two phases. At $T_1\lesssim T_1^*$, $\la S \ra$ is much smaller than the value expected for random vectors in the Hilbert space, $S_{\rm Th}\approx L/2 \ln2$~\cite{Page93}, which signals ergodicity breaking. Moreover, at $T_1\lesssim 0.6$ the entanglement entropy grows very weakly with system size, consistent with area-law in 1D. On the contrary, at large $T_1>T_1^*$, $\la S \ra$ approaches $S_{\rm Th}$, indicating that almost all eigenstates are essentially random vectors in the Hilbert space, as expected in the ergodic phase. 

It is also instructive to study the fluctuations of entanglement entropy, as they have been shown to provide a useful probe of the MBL-delocalization transition in time-independent models~\cite{Kjall14}. The disorder-averaged fluctuations of $S$, defined as $\Delta S =\sqrt{\la (S-\la S \ra)^2 \ra}$ are expected to be small deeply in the delocalized phase, as well as in the MBL phase: in the former case, almost all eigenstates are highly entangled, with $S\approx S_{\rm Th}$, with small fluctuations around this value, while in the latter case, $S$ obeys area-law and is therefore small, as are its state-to-state fluctuations. In contrast, at the transition $S$ has a broad distribution~\cite{Serbyn13-2,Kjall14}, and therefore its fluctuations are maximal. Thus, the localization-delocalization transition can be detected by the location of the peak in $\Delta S$. Fig.~\ref{Fig2}(inset) shows $\Delta S$ as a function of $T_1$. Entanglement fluctuations $\Delta S$ exhibit a maximum at $T_1\approx 1.1$ that roughly agrees with $T_1^*$ value found from analyzing level statistics; further, we observe a slight drift of the maximum with the system size, similar to the previous study of the static case~\cite{Kjall14}. We attribute the difference between the position of the maximum in $\Delta S$ and value $T_1$ determined from the level statistics, to the finite-size effects. We have also directly tested the ETH and its violation in the MBL phase in the Floquet eigenstates, finding behaviour consistent with the existence of two phases~\cite{supp} (see also Ref.~\cite{Kim_ETH}, where ETH for driven ergodic systems was tested).

\begin{figure}[htb]
\begin{center}
\includegraphics[width=\columnwidth]{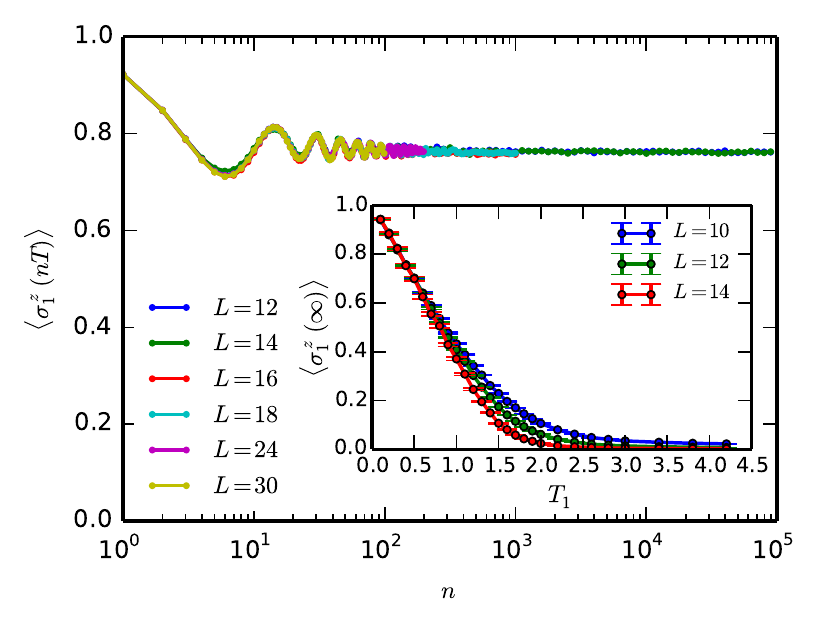}
\caption{ \label{Fig3} Dynamical properties: decay of magnetization at a given site $I=1$ for a N\'eel initial configuration. Inset: Long-time magnetization remains non-zero in the MBL phase as the system size is increased. In the delocalized phase, magnetization decays to zero at long times. Averaging was performed over 6000 disorder realizations.}
\end{center}
\end{figure}
{\bf Dynamics.} We next study the dynamical properties of the model (\ref{eq:H},\ref{eq:kick}). We consider a standard quantum quench protocol: the system is initially prepared in a N\'eel (product) state $|\psi_0\ra$ of spins $\sigma_i^z=\pm 1$ at $t=0$, and this state is evolved under the Hamiltonian (\ref{eq:H},\ref{eq:kick}) at $t>0$. This protocol is particularly easy to simulate using Krylov subspace projection methods~\cite{expokit} or time-evolving block decimation~\cite{tebd} method, both of which allow us to access larger systems beyond ED due to the sufficiently slow growth of entanglement in the MBL phase. For the TEBD algorithm we use a second order Trotter decomposition with time step $\Delta t=0.1$.  The growth of the bond dimension is controlled by requiring the neglected weight to be less than $10^{-7}$ at each Schmidt decomposition.

{\it Local observables.} We first focus on the evolution of local observables, and compute the expectation value of the spin on a given site $I$, $\sigma_{I}^z(t)$, and its long-time limit $\la \sigma_I^z (\infty) \ra$~\cite{Vasseur14,Chandran14,Serbyn14}. Fig.\ref{Fig3} illustrates the time evolution $\sigma_{I}^z(t)$ for the N\'eel initial state $|\psi_0\ra$ and site $I=1$, and for system sizes ranging from $L=10-14$ (obtained via ED), $L=16,18$ obtained using Krylov subspace projection, and $L=24,30$ obtained using TEBD. We find that the on-site magnetization remains finite at very long times even for the largest systems without any visible finite-size effects. This indicates that the MBL phase remains stable in the thermodynamic limit.

The long-time average $\la \sigma_I^z (\infty) \ra$ can be expressed in terms of the Floquet eigenstates as  
$\la \sigma_I^z (\infty) \ra=\lim_{t\to \infty}\frac{1}{t}\int_0^t \la \psi_0| \sigma_I^z (t') |\psi_o \ra \, dt'$, which in terms of the eigenstates $|\psi_\alpha\ra$ reads $\sum_\alpha \la \psi_\alpha| \sigma_{I}^z|\psi_\alpha\ra |\la\psi_0|\psi_\alpha \ra|^2$. The long-time value $\la \sigma_I^z (\infty) \ra$, calculated using ED, and averaged over 6000 disorder realizations, is illustrated in Fig.~\ref{Fig3}(inset). This quantity behaves differently in the two phases: at $T\lesssim T_1^*$, $\la \sigma_I^z (\infty) \ra$ is positive and weakly dependent on the system size, which shows that in the MBL phase the local memory of the initial state is retained.  Deep in the ergodic phase, at $T_1\gg T_1^*$, $\la \sigma_I^z (\infty) \ra\to 0$, reflecting the decay of the initial magnetization and therefore a loss of the memory of the initial state. 

\begin{figure}[t]
\begin{center}
\includegraphics[width=1\columnwidth]{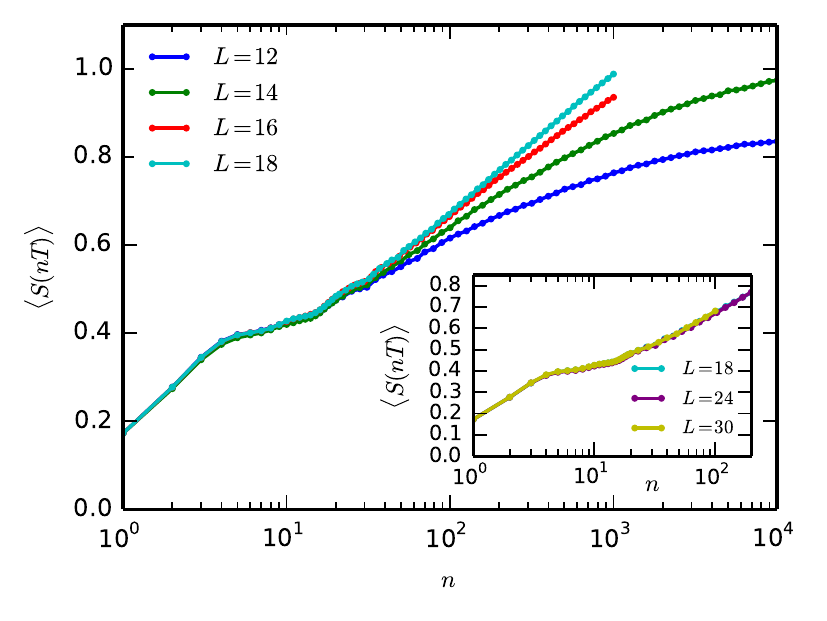}
\caption{ \label{Fig4} Disorder-averaged entanglement entropy following a quantum quench, for the N\'eel initial state. Data for  system sizes $L=12,14$ was obtained by ED, for $L=16,18$ using Krylov subspace projection, and $L=24,30$ using TEBD. Averaging performed over 6000 disorder realizations.}
\end{center}
\end{figure}

{\it Entanglement growth.} Finally, we explored the spreading of entanglement following a quantum quench, known to be a sensitive probe of many-body localization: in the MBL phase, entanglement grows logarithmically in time~\cite{Moore12,Serbyn13-1,Serbyn13-2,Huse13,Prosen08}, while in the ergodic phase, as well as in Bethe-ansatz-integrable systems, it grows linearly in time~\cite{Calabrese05,Chiara05,Kim13}. The disorder-averaged entanglement entropy as a function of time, calculated for fixed $T_1=0.4$ and the symmetric bipartition, is shown in Fig.~\ref{Fig4}. Averaging was performed over 6000 disorder realizations. Entanglement initially rises from zero, followed by a plateau and a logarithmic growth for several decades in time, $\la S(t)\ra\propto \ln (t)$. This behavior is qualitatively similar to that found in the MBL phase in systems with time-independent Hamiltonians~\cite{Serbyn13-1,Serbyn13-2,Moore12,Huse13}, which gives further support for the existence of the MBL phase in driven systems with strong disorder.  

{\bf Local integrals of motion and effective description of the driven MBL phase.} In order to understand the spectral and dynamical properties of the MBL phase observed in the numerical simulations, we propose that this phase is characterized by an extensive number of local integrals of motion~\cite{Serbyn13-2,Huse13}. First, we note that the area-law entanglement of the Floquet eigenstates suggests that they can be obtained from the product states (in the $\sigma_{i}^z=\pm 1$ basis) by a quasi-local unitary transformation $U$ which brings the Floquet operator into a diagonal form in that basis: 
$U\hat{F}U^\dagger=\hat{F}_{\rm diag}$. Since $L$ of the operators $\sigma_i^z$ commute with $\hat{F}_{\rm diag}$, we can introduce a set of $L$ ``pseudospin" operators $\tau_i^z=U^\dagger \sigma_i^z U$. These operators commute with the Floquet operator $[\hat{F},\tau_i^z]=0$, as well as with each other
$[\tau_i^z,\tau_j^z]=0$. 
Operators $\tau_i^z$ have eigenvalues $\pm 1$ and therefore satisfy the relation $(\tau_i^z)^2=1$; they can be viewed as  $z$-components of some ``effective" spins. We emphasize that the operators $\tau_i^z$ can be introduced for any driven system, but the special property of the MBL phase is that their support is localized near site $i$, and they affect remote physical degrees of freedom exponentially weakly.  In terms of $\tau$-operators, the operator $F$ takes a simple form, as it can only depend on $\tau_i^z$ operators and their products (but not on the $\tau_i^x,\tau_i^y$ operators). It is convenient to represent $\hat{F}$ as 
\be\label{eq:Feff}
\hat{F}=e^{-iH_{\rm eff}(\{\tau_i^z \})},
\ee
 where $H_{\rm eff}(\{\tau_i^z \})$ is a real function of operators $\tau_i^z$. (Such a representation takes into account the fact that eigenvalues of $\hat{F}$ have absolute value one). Further, since $(\tau_i^z)^2=1$, $H_{\rm eff}$ can generally be written as
\be\label{eq:Heff}
H_{\rm eff}(\{\tau_i^z \})=\sum_{i} \tilde h_i \tau_i^z+\sum_{ij} J_{ij} \tau_i^z \tau_j^z+\sum_{ijk} J_{ijk} \tau_i^z\tau_j^z\tau_k^z + \ldots
\ee
It is natural to assume that in the MBL phase the couplings $J$ between remote effective spins decay exponentially with distance, similar to the static case~\cite{Serbyn13-2,Huse13}; we note that long-range interactions, in particular, would be inconsistent with Lieb-Robinson bounds on information propagation~\cite{LiebRobinson} satisfied by the operator $\hat{F}$. 

The effective model introduced above naturally explains the spectral and dynamical properties of the MBL phase established numerically, e.g., the absence of decay of the on-site magnetization at long times and the logarithmic growth of entanglement, which directly follows from Eqs.(\ref{eq:Feff},\ref{eq:Heff}) and exponential decay of interactions between remote effective spins~\cite{Serbyn13-1,Serbyn13-2,Huse13}. To provide further justification for the effective description (\ref{eq:Feff},\ref{eq:Heff}), we have also numerically constructed~\cite{supp} the local integrals of motion following Ref.~\cite{Chandran14}. These form an extensive set, although they are not identical to $\tau_i^z$ operators. 

{\bf Discussion.} We have demonstrated the existence of two dynamical regimes in periodically driven systems described by local interacting Hamiltonians with quenched disorder. In particular, we have identified a many-body localized phase, in which ergodicity is broken. We argued that the MBL phase is characterized by extensively many emergent, quasi-local conservation laws. This implies that the dynamics of Floquet MBL systems is described by an effective quasi-local {\it time-independent} Hamiltonian $H_{\rm eff}$, which is itself many-body-localized. This is in sharp contrast to the ergodic phase, where the Floquet Hamiltonian does not have a quasi-local representation~\cite{Alessio14,Lazarides14,Ponte14}. An interesting open question is whether the Magnus expansion~\cite{magnus} converges in the MBL phase. 

Another implication of our results is that MBL does not rely on global conservation laws. Further, MBL phase is robust under sufficiently weak periodic driving, and  there exists a finite driving threshold above which transport is restored, and the system ultimately delocalizes. This may serve as an experimental signature of the many-body localization. An interesting subject for future research, relevant for experiments in disordered solid-state systems, is to study periodically driven MBL system weakly coupled to a thermal bath (we note that spectral properties of a static MBL system coupled to a bath were recently considered in Refs.~\cite{Nandkishore14, Johri14}).

{\bf Acknowledgements.} We thank Anushya Chandran, David Huse, Isaac Kim, and Yuan Wan for enlightening discussions. We acknowledge support by Alfred Sloan Foundation and Early Researcher Award by the Government of Ontario (DA). Research at Perimeter Institute is supported by the Government of Canada through Industry Canada and by the Province of Ontario through the Ministry of Economic Development \& Innovation. 

{\it Note added.} During the completion of this manuscript, we became aware of a related recent work~\cite{Lazarides14-2}.

\newpage

\vspace{0.1cm}
\begin{center}
\begin{widetext}
{\large \bf Supplemental Online Material for ``Many-body localization in periodically driven systems"}
\\
\vspace{0.5cm}
Pedro Ponte, Z. Papi\'c, Fran\c{c}ois Huveneers, and Dmitry A. Abanin
\end{widetext}
\end{center}

\vspace{0.5cm}

Here we provide direct numerical tests of the eigenstate thermalization hypothesis (ETH) in driven MBL and delocalized phases. We also explicitly construct local integrals of motion in the MBL phase for the Floquet problem using the method of Ref.~\cite{Chandran14}. 

\section{Testing the ETH}

According to the ETH, in the delocalized phase the expectation value of a local operator $\mathcal{O}$ in all Floquet eigenstates should converge, in the thermodynamic limit, to the prediction of the canonical ensemble with infinite temperature  
$$\mathcal{O}_{\infty} =\frac{1}{\mathcal{D}}\mathrm{Tr }\, \mathcal{O},$$ 
where $\cal{D}$ is the Hilbert space dimension. We test ETH and its violation in the MBL phase by examining the deviation of the expectation value of $\mathcal{O}$ in individual eigenstates, $\la\mathcal{O} \ra_\alpha$ from $\mathcal{O}_{\infty}$: 
$$\Delta \mathcal{O}=\langle |\langle \mathcal{O} \rangle_\alpha - \mathcal{O}_\infty|\rangle,$$
where averaging is performed over all Floquet eigenstates for each disorder realization, and then over different disorder realization. We expect this quantity to approach zero in the delocalized phase, as $L\to \infty$. On the other hand, in the MBL phase this quantity should remain finite as we extrapolate the chain size $L$ to infinity. 
\begin{figure}[htb]
\begin{center}
\includegraphics[width=0.49\columnwidth]{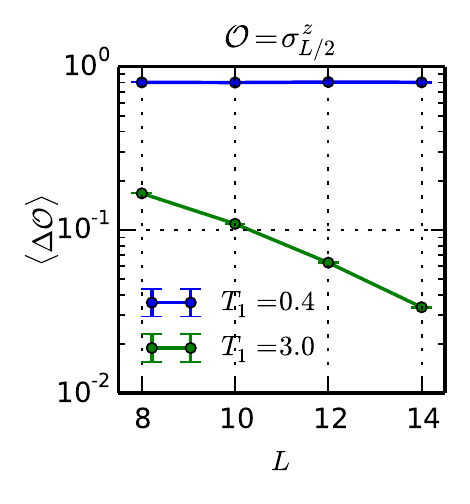}
\includegraphics[width=0.49\columnwidth]{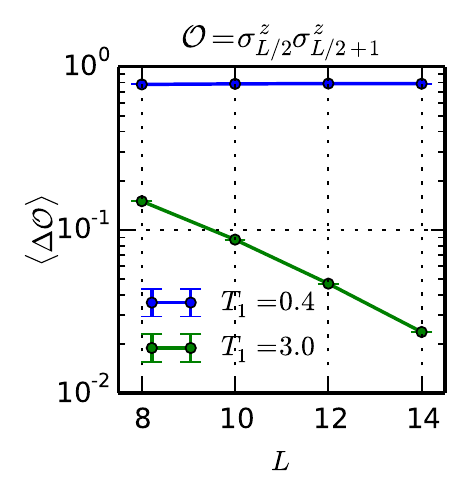}
\includegraphics[width=0.49\columnwidth]{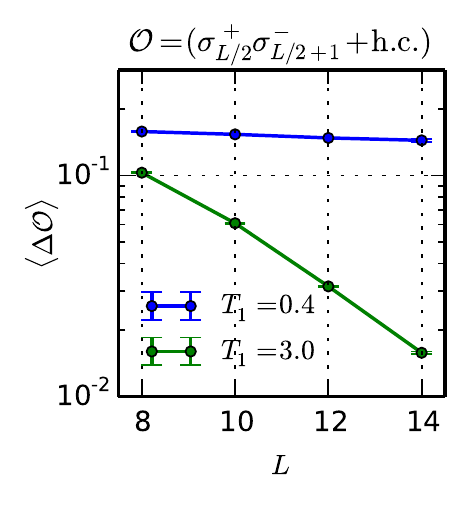}
\includegraphics[width=0.49\columnwidth]{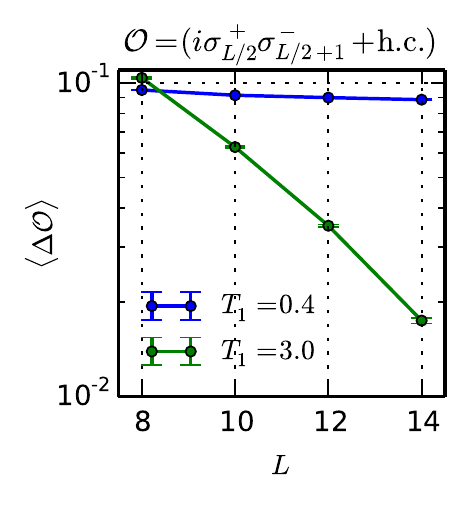}
\caption{ \label{FigA1} Deviation of expectation values of local operators from their infinite-temperature values given by the canonical ensemble. Different plots correspond to various choices of $S_z$-preserving local operators $\mathcal{O}$ acting on the sites in the middle of the chain, and $T_1=0.4, 3.0$ correspond to MBL and delocalized phases, respectively.}
\end{center}
\end{figure}

In Fig.~\ref{FigA1} we show numerical results for local operators $\mathcal{O}$ that act on the two neighbouring sites in the middle of the chain and conserve $S^z$. We studied the following four operators: 
\begin{eqnarray}
\nonumber && \mathcal{O}_1= \sigma^z_{L/2}, \\
\nonumber && \mathcal{O}_2= \sigma^z_{L/2}\sigma^z_{L/2+1}, \\
\nonumber && \mathcal{O}_3= (\sigma^+_{L/2}\sigma^-_{L/2+1}+\sigma^-_{L/2}\sigma^+_{L/2+1}), \\
\nonumber && \mathcal{O}_4= i(\sigma^+_{L/2}\sigma^-_{L/2+1}-\sigma^-_{L/2}\sigma^+_{L/2+1}).
\end{eqnarray}
As expected, in the MBL phase ($T_1=0.4$), $\langle\Delta \mathcal{O}\rangle$ changes weakly with system size suggesting that it remains finite in the thermodynamic limit. In the delocalized phase ($T_1=3.0$), $\Delta \mathcal{O}$ approaches zero with increasing system size suggesting that each Floquet eigenstate behaves as an infinite temperature thermal state for local observables in the thermodynamic limit. We note that all four operators show nearly identical behaviour. Thus the direct test of the ETH is consistent with the presence of two phases with markedly different properties of their eigenstates.

\section{Local integrals of motion}

\begin{figure*}[htb]
\begin{minipage}[l]{0.45\linewidth}
\centering
\includegraphics[width=0.95\textwidth]{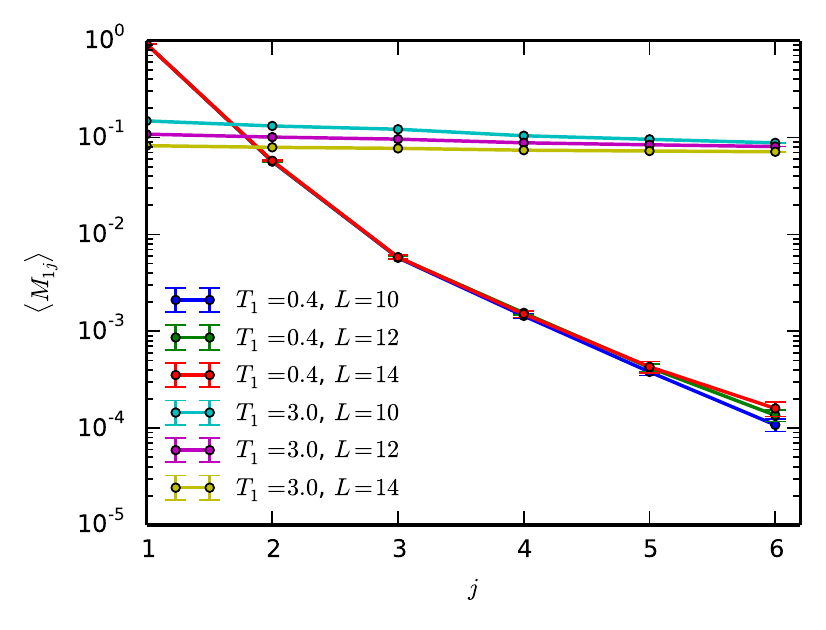}
\label{FigA2a}
\end{minipage}
\hspace{0.2cm}
\begin{minipage}[l]{0.45\linewidth}
\centering
\includegraphics[width=0.95\textwidth]{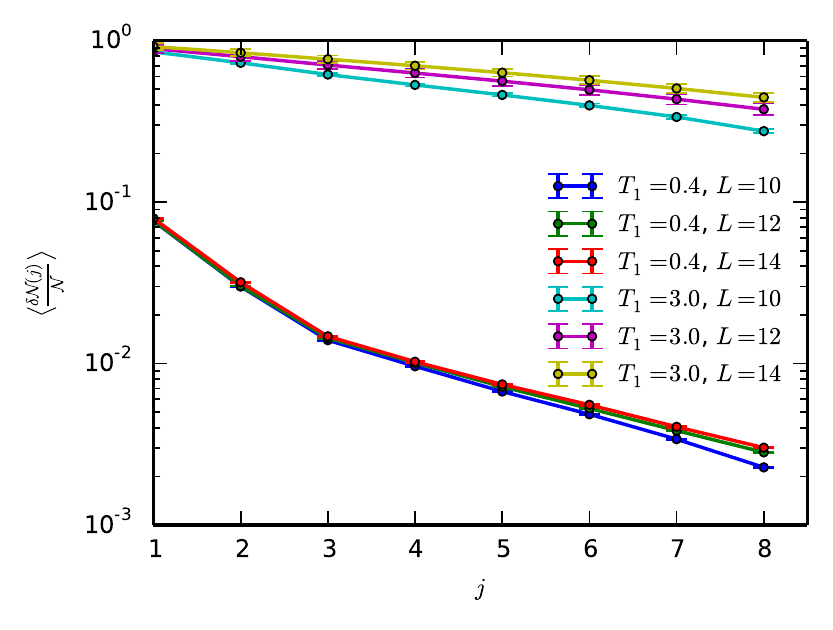}
\label{FigA2b}
\end{minipage}
\caption{Local integrals of motion in the MBL and delocalized phase. (Left) Median magnetization $M_{1j}$ as a function of distance $|1-j|$. (Right) Median difference between the total norm $\mathcal{N}$, and partial norm $\delta \mathcal{N}(j)$, divided by $\mathcal{N}$. The exponential decay of this quantity with distance $|j-1|$ demonstrates that in the MBL phase ($T_1=0.4$) operator $\bar\sigma_1^z$ is a quasi-local integral of motion. In the delocalized phase ($T_1=3$), this operator becomes non-local.}
\label{FigA2}
\end{figure*}

To further support the effective model of the MBL phase introduced in the main text, we now explicitly construct an extensive set of quasi-local integrals of motion. Following Ref.~\cite{Chandran14}, we consider infinite-time average of the operator $\sigma^z_1$, denoted by  $\bar\sigma^z_1$, which is always an integral of motion. We now demonstrate that in the MBL phase this operator is a {\it quasi-local} integral of motion. 

First, we note that operator $\bar\sigma^z_1$ describes the spreading of magnetization, initially prepared on site 1. To illustrate this, consider the infinite-temperature ensemble, in which spin 1 was initially prepared in the up-state, described by the density matrix 
$$
\rho(0)=2^{-L} (1+\sigma_1^z) \otimes \mathds{1}_{i\neq 1}.
$$
this initial magnetization will spread over the chain with time.  Upon time averaging, the density matrix can be expressed in terms of the operator $\bar\sigma^z_1$ as follows:
\begin{eqnarray}
\nonumber \bar{\rho}=2^{-L}(1+\bar{\sigma}_1^z), 
\end{eqnarray}
where $\bar{\sigma}_1^z=\lim_{T\to\infty} \frac{1}{T}\int_0^T \sigma_1(z)(t)dt$. Thus the spreading of the initial magnetization over the chain at long times can be related to the properties of operator $\bar\sigma^z_1$. The long-time  magnetization on site $j$ is given by: 
$$
M_{1j}=\mathrm{Tr} (\bar{\rho}\sigma_j^z)=\frac{1}{2^L}\mathrm{Tr}\left(\bar{\sigma}^z_1\sigma^z_j\right). 
$$
Fig.~\ref{FigA2} illustrates that $M_{11}$ is on the order of unity, but $M_{1j}$ decays over several orders of magnitude as a function of $|j-1|$ in the MBL phase ($T_1=0.4$). This is consistent with $\bar{\sigma}^z_1$ being a quasi-local operator. Conversely, in the delocalized regime ($T_1=3$), the magnetization is nearly uniformly spread over all sites $j$ and has a stronger dependence on the chain size $L$, thus, operator $\bar{\sigma}^z_1$ becomes non-local. 

To further test the quasi-locality of operator $\bar\sigma_1^z$ in the MBL phase, we examined the partial norm
$$\mathcal{N}(j)=\frac{1}{2^{j}}\textrm{Tr} (\bar{\sigma}^A \bar{\sigma}^A ),
$$ 
where  
$$\bar{\sigma}^A \equiv \frac{1}{2^{L-j}} {\textrm{Tr}}_{\bar{A}} \bar{\sigma}_1^z,$$
and $A$ is the region containing sites 1 to $j$ and $\bar{A}$ its complement. In Fig.~\ref{FigA2} we illustrate the normalized difference $\frac{\delta \mathcal{N}(j)}{\mathcal{N}}=\frac{\mathcal{N}-\mathcal{N}(j)}{\mathcal{N}}$ vs $j$, where $\mathcal{N}=\frac{1}{2^L} \mathrm{Tr} \bar{\sigma}^z_1 \bar{\sigma}^z_1$ is the total norm of the operator $\bar{\sigma}^z_1$. This quantity describes how well the operator $\bar{\sigma}_1^z$ can be approximated by operators with a finite support, and therefore tests whether this operator is quasi-local. It is evident from Fig.~\ref{FigA2} that $\frac{\delta \mathcal{N}(j)}{\mathcal{N}}$ approaches zero exponentially in distance $|j-1|$ in the MBL phase, indicating that the operator $\bar{\sigma}^z_1$ is indeed a quasi-local integral of motion. We note that similar quasi-local integrals of motion, $\bar{\sigma}^z_i$, can be constructed for other sites, $i=2,..,L$, and they form an extensive set of LIOMs.


\begin{thebibliography}{16}

\expandafter\ifx\csname natexlab\endcsname\relax\def\natexlab#1{#1}\fi
\expandafter\ifx\csname bibnamefont\endcsname\relax
  \def\bibnamefont#1{#1}\fi
\expandafter\ifx\csname bibfnamefont\endcsname\relax
  \def\bibfnamefont#1{#1}\fi
\expandafter\ifx\csname citenamefont\endcsname\relax
  \def\citenamefont#1{#1}\fi
\expandafter\ifx\csname url\endcsname\relax
  \def\url#1{\texttt{#1}}\fi
\expandafter\ifx\csname urlprefix\endcsname\relax\def\urlprefix{URL }\fi
\providecommand{\bibinfo}[2]{#2}
\providecommand{\eprint}[2][]{\url{#2}}

\bibitem{Polkovnikov11} A. Polkovnikov, K. Sengupta, A. Silva, and M. Vengalattore, Rev. Mod. Phys. {\bf 83}, 863 (2011). 

\bibitem{Bloch08} I. Bloch, J. Dalibard, and W. Zwerger, Rev. Mod. Phys. {\bf 80}, 885 (2008).

\bibitem{deutsch} J. M. Deutsch, Phys. Rev. A {\bf 43}, 2046 (1991).
\bibitem{srednicki} M. Srednicki, Phys. Rev. E {\bf 50}, 888 (1994).

\bibitem{Rigol08} M. Rigol, V. Dunjko, and M. Olshanii, Nature {\bf 452},854 (2008). 


\bibitem{Basko06} D. Basko, I. Aleiner, and B. Altshuler, Annals of Physics {\bf 321}, 1126 (2006). 

\bibitem{Mirlin05} I. V. Gornyi, A. D. Mirlin, and D. G. Polyakov, Phys. Rev. Lett. {\bf 95}, 206603 (2005). 

\bibitem{Oganesyan07} V. Oganesyan and D. A. Huse, Phys. Rev. B {\bf 75}, 155111 (2007). 

\bibitem{Prosen08} M. Znidaric, T. Prosen, and P. Prelovsek, Phys. Rev. B {\bf 77}, 064426 (2008). 


\bibitem{Pal10} A. Pal and D. A. Huse, Phys. Rev. B {\bf 82}, 174411 (2010). 

\bibitem{Moore12} J. H. Bardarson, F. Pollmann, and J. E. Moore, Phys. Rev. Lett. {\bf 109}, 017202 (2012). 

\bibitem{Serbyn13-1} M. Serbyn, Z. Papi\'c, and D. A. Abanin, Phys. Rev. Lett. {\bf 110}, 260601 (2013). 

\bibitem{Serbyn13-2} M. Serbyn, Z. Papi\'c, and D. A. Abanin, Phys. Rev. Lett. {\bf 111}, 127201 (2013). 

\bibitem{Huse13} D. A. Huse, R. Nandkishore, and V. Oganesyan, arxiv:1408.4297.

\bibitem{Vosk13} R. Vosk and E. Altman, Phys. Rev. Lett. {\bf 110}, 067204 (2013). 

\bibitem{bauer} B. Bauer and C. Nayak, J. Stat. Mech. (2013) P09005.

\bibitem{Pekker14} D. Pekker, G. Refael, E. Altman, E. Demler, and V. Oganesyan, Phys. Rev. X {\bf 4}, 011052 (2014). 

\bibitem{Calabrese05}
P. Calabrese and J. Cardy, J. Stat. Mech. P04010 (2005). 

\bibitem{Chiara05}
G. De Chiara, S. Montangero, P. Calabrese, and R. Fazio, J. Stat. Mech. P03001 (2006). 

\bibitem{Kim13} H. Kim and D. A. Huse, Phys. Rev. Lett. {\bf 111}, 127205 (2013). 

\bibitem{Chandran14} A. Chandran, I. H. Kim, G. Vidal, and D. A. Abanin, arXiv:1407.8480 (2014). 

\bibitem{Ros14} V. Ros, M. Mueller, and A. Scardicchio, arXiv:1406.2175 (2014). 


\bibitem{Prosen98} T. Prosen, Phys. Rev. Lett. {\bf 80}, 1808 (1998). 

\bibitem{Alessio13} L. D'Alessio and A. Polkovnikov, Annals of Physics {\bf 333}, 19 (2013). 

\bibitem{Alessio14} L. D'Alessio and M. Rigol, arXiv:1402.5141 (2014). 

\bibitem{Lazarides14} A. Lazarides, A. Das, and R. Moessner, Phys. Rev. E {\bf 90}, 012110 (2014). 

\bibitem{Ponte14} P. Ponte, A. Chandran, Z. Papi\'c, and D. A. Abanin, arXiv:1403.6480 (2014). 


\bibitem{casati} G. Casati and J. Ford, \emph{Stochastic Behavior in Classical and Quantum Hamiltonian Systems}, 1st ed., Vol. {\bf 93} (Springer Berlin Heidelberg, 1979).

\bibitem{grempel} D. R. Grempel, R. E. Prange, and S. Fishman, Phys. Rev. A {\bf 29}, 1639 (1984).

\bibitem{flmoore} F. L. Moore, J. C. Robinson, C. Bharucha, P. E. Williams, and M. G. Raizen, Phys. Rev. Lett. {\bf 73}, 2974 (1994).

\bibitem{lemarie} G. Lemari\'e, J. Chab\'e, P. Szriftgiser, J. C. Garreau, B. Gr\'emaud, and D. Delande, Phys. Rev. A {\bf 80}, 043626 (2009).


\bibitem{haake} Fritz Haake, \emph{Quantum signatures of chaos}, Springer-Verlag, Berlin, 2nd
edition (1991).

\bibitem{izrailev} F. M. Izrailev and D. L. Shepelyanskii, Theor. Math. Phys. {\bf 43}, 1586 (1980).

\bibitem{dittrich} T. Dittrich and R. Graham, Annals Phys. {\bf 200}, 363 (1990).

\bibitem{Page93}
D. N. Page, Phys. Rev. Lett. {\bf 71}, 1291 (1993). 



\bibitem{Kjall14} J. A. Kj\"all, J. H. Bardarson, and F. Pollmann, Phys. Rev. Lett. {\bf 113}, 107204 (2014).


\bibitem{supp} See Supplemental Online Material.


\bibitem{Kim_ETH} H. Kim, T. N. Ikeda, and D. A. Huse, arXiv:1408.0535 (2014). 


\bibitem{expokit} R. B. Sidje, ACM Trans. Math. Softw., {\bf 24}, 130 (1998).



\bibitem{tebd} G. Vidal, Phys. Rev. Lett. {\bf 91}, 147902 (2003).

\bibitem{Vasseur14} R. Vasseur, S. A. Parameswaran, and J. E. Moore, arXiv:1407.4476 (2014). 

\bibitem{Serbyn14} M. Serbyn, Z. Papi\'c, and D. A. Abanin, arXiv:1408.4105 (2014). 

\bibitem{LiebRobinson} E. H. Lieb and D. Robinson, Commun. Math. Phys. {\bf 28}, 251 (1972). 


\bibitem{magnus} W. Magnus, Pure Appl. Math {\bf 7}, 649 (1954).


\bibitem{Nandkishore14}
R. Nandkishore, S. Gopalakrishnan, and D. A. Huse, Phys. Rev. B {\bf 90}, 064203 (2014).

\bibitem{Johri14} S. Johri, R. Nandkishore, and R. N. Bhatt, arXiv:1405.5515 (2014).
   
\bibitem{Lazarides14-2} A. Lazarides, A. Das, R. Moessner, arXiv:1410.3455 (2014). 



\end{thebibliography}
\end{document}